\documentclass[11pt]{article}
\usepackage{amssymb,latexsym,amsmath,amsbsy,amsthm}
\usepackage[dvips]{graphicx}
\usepackage{xcolor}
\usepackage{cite}
\usepackage{stmaryrd}  
\def\deg{\mathop{\rm deg}\nolimits}

\def\N{{\mathbb N}}

\def\gl{\mathfrak{gl}}
\def\ssl{\mathfrak{sl}}
\def\so{\mathfrak{so}}
\def\osp{\mathfrak{osp}}

\def\t{\theta}
\newtheorem{theo}{Theorem}
\newtheorem{defi}{Definition}

\begin{document}
\begin{center}
{\Large \bf
Generalized boson and fermion operators \\[2mm] 
with a maximal total occupation property} \\[5mm]
{\bf N.I.~Stoilova}\footnote{Corresponding author}\\[1mm] 
Institute for Nuclear Research and Nuclear Energy, Bulgarian Academy of Sciencies,\\ 
Boul.\ Tsarigradsko Chaussee 72, 1784 Sofia, Bulgaria\\[2mm] 
{\bf J.\ Van der Jeugt}\\[1mm]
Department of Applied Mathematics, Computer Science and Statistics, Ghent University,\\
Krijgslaan 281-S9, B-9000 Gent, Belgium\\[2mm]
E-mail: stoilova@inrne.bas.bg, JorisVanderJeugt@ugent.be
\end{center}

\vskip 2 cm

\begin{abstract}
\noindent 
We propose a new generalization of the standard (anti-)commutation relations for creation and annihilation operators of bosons and fermions.
These relations preserve the usual symmetry properties of bosons and fermions.
Only the standard (anti-)commutator relation involving one creation and one annihilation operator is deformed by introducing fractional coefficients, 
containing a positive integer parameter~$p$.
The Fock space is determined by the classical definition.
The new relations are chosen in such a way that the total occupation number in the system has the maximum value~$p$.
From the actions of creation and annihilation operators in the Fock space, 
a group theoretical framework is determined, and from here the correspondence with known particle statistics is established.
\end{abstract}

\vskip 10mm
\noindent Generalized bosons and fermions 


\setcounter{equation}{0}
\section{Introduction} \label{sec:A}%

Generalizations or deformations of canonical quantum statistics have a long history~\cite{Gentile}.
These generalizations are often intermediate between Bose-Einstein statistics (bosons) and Fermi-Dirac statistics (fermions), and therefore referred to as ``intermediate statistics''.
Various types of generalizations have been introduced and discussed: in the context of quantum field theory,
condensed matter physics, quantum groups or deformed 
algebras~\cite{Wigner,Green,Greenberg,Wilczek, Wilczek2,Haldane,Wu,Khare,Pusz,Macfarlane,Feng,Bonatsos,Curtright,Polychronakos,Mishra, Tichy, Dai2022}.
The most successful generalization is fractional statistics in two dimensions, as it has been observed experimentally~\cite{Camino}.

Interesting generalizations arise from deforming the relations between (boson or fermion) creation and annihilation operators, 
see e.g.~\cite{Trifonov2009,Trifonov2012,DaiXie,Borasi,Sanchez,WangHazzard} and references therein.
A lot of attention has gone to so-called parastatistics.
Herein, the relations between creation and annihilation operators are triple relations (instead of quadratic), 
and for the corresponding Fock space an ``order of statistics'' parameter $p$ is introduced~\cite{Green,Greenberg}.
Parabosons and parafermions have a natural group theoretical setting: 
parabosons correspond to the Lie superalgebra $\osp(1|2n)$~\cite{Ganchev} with Fock space its 
unitary irreducible representation with lowest weight $(\frac{p}{2}, \frac{p}{2},\ldots, \frac{p}{2})$~\cite{FockPB}, and parafermions to the Lie algebra $\so(2m+1)$~\cite{KR} with Fock space its unitary irreducible representation with lowest weight $(-\frac{p}{2}, -\frac{p}{2},\ldots, -\frac{p}{2})$~\cite{FockPF}. 
Palev~\cite{Palev_HT,Palev_97,Palev1980A,Palev1982} and collaborators~\cite{Palev_J2002,Jellal,Palev2003,SJLA,SJLSA} generalized this idea to other Lie (super)algebras, 
leading to so-called $A, B, C, D$-statistics or $A, B, C, D$-superstatistics.
In~\cite{Palev_97,Palev1980A,Palev_J2002,Jellal,Palev2003}, 
the emphasis is on statistics of type $A$:
the relations between creation and annihilation operators are not quadratic but triple relations (following from a Lie algebraic approach), 
and the corresponding Fock space is determined by a seemingly unnatural condition (involving the so-called order of statistics).
Although the statistical properties are interesting, the starting point is hard to motivate from a physical point of view.
Our current results are in this context, following however a completely different approach.

The generalized system of fermions or bosons considered in this paper is as close as possible to ordinary fermions and bosons.
The only difference is that we require that the total number of particles of a state in the system is bounded by some positive integer~$p$.
This is accomplished by a minor modification of the (anti-)commutation relations between creation and annihilation operators.
The relations proposed here are still quadratic in the creation and annihilation operators. 
For example, for fermions the creation operators $f^+_i$ anti-commute among themselves, and also the annihilation operators $f^-_i$ anti-commute.
Only the anti-commutator $f_i^- f_j^+ + f_j^+ f_i^- = \delta_{ij}$ is modified by introducing fractional coefficients in this relation, involving the parameter~$p$.
For bosons, the modification is similar.
The Fock space is determined by the same conditions as for ordinary fermions or bosons: 
it is generated by a single vacuum state that is annihilated by all annihilation operators.
The physical context can be of any kind: we speak of particle creation and annihilation operators (where particle could also mean quasiparticle, excitation, etc.),
and the index $i$ (such as in $f^+_i$ or $f^-_i$) can refer to an internal index or an orbital, for example.

It is very satisfying that these new quadratic relations only allow the construction of a complete set of basis vectors for the Fock space. 
From these Fock spaces, the basic required properties are again transparent:
for the generalized fermions, each orbital is occupied by at most one particle (exclusion principle), 
and on top of that the total number of particles in the system is restricted by~$p$;
for the generalized bosons, there is no restriction for the number of particles on each orbital, 
but the total number of particles in the system is again restricted by~$p$.

Once the analysis of the Fock space has been accomplished, one can continue and give a group theoretical interpretation of this space and the actions of the operators.
At that moment, it becomes apparent that the Fock space constructed here coincides with the ones of~\cite{Jellal} or~\cite{Palev2003}.
In other words, the quantum statistics appearing in this paper is not new.
But let us emphasize again that the approach of this paper is totally different and novel: 
it matches the classical approach of the algebra of ordinary fermion and boson creation and annihilation operators.

The structure of the paper is as follows: in Section~\ref{sec:B} we propose the new anti-commutation relations for the generalized fermions.
From the mathematical point of view, we are then dealing with an algebra ${\cal F}$ generated by the creation and annihilation operators and characterized by quadratic relations therein.
The Fock space for this algebra is defined in the standard way, and the new relations allow us to construct a simple basis for the Fock space and for the action of the generators.
In Section~\ref{sec:C} we show that there is an underlying group theoretical framework: 
the generators (in case there is a finite number of them) can be identified as certain generators of the Lie superalgebra $\gl(1|n)$,
and the Fock space is then a certain irreducible representation of $\gl(1|n)$.
In the following sections~\ref{sec:D} and~\ref{sec:E}, we perform the same analysis for generalized bosons.
Here, the Fock space can be identified as an irreducible representation of the Lie algebra $\gl(1+n)$.
Some final remarks, and a toy model example, are given in a concluding section~\ref{sec:F}.

\section{Generalized fermion operators}
\setcounter{equation}{0} \label{sec:B}

Consider a set of ordinary fermion creation and annihilation operators $F_i^\pm$, where $i$ belongs to some index set $I$. 
This set $I$ can be finite ($I=\{1,\ldots,n\}$) or infinite ($I=\N$). 
The defining relations are:
\begin{align}
& \{ F_i^+,F_j^+\} = \{ F_i^-,F_j^-\} = 0, \label{FpFp}\\
& \{ F_i^-,F_j^+\} = \delta_{ij}, \qquad (i,j\in I). \label{FmFp}
\end{align}
The representation space (Fock space) is generated by a vacuum vector $|0\rangle$ and determined by
\begin{equation}
F_i^- |0\rangle =0\qquad (i\in I),  \qquad \langle 0 | 0\rangle = 1
\end{equation}
and the hermiticity conditions $(F_i^\pm)^\dagger = F_i^\mp$.
The orthonormal basis vectors of the Fock space can be written as
\begin{equation}
|\t\rangle = |\t_1,\t_2,\ldots\rangle= (F_1^+)^{\t_1} (F_2^+)^{\t_2} \ldots |0\rangle,
\qquad \t_i\in\{ 0,1\}, \qquad \sum_i\t_i<\infty .
\end{equation}
For $I=\{1,\ldots,n\}$ it is well known that the Fock space has dimension $2^n$.
The action of the creation and annihilation operators on the above vectors is given by
\begin{align}
 F_i^- |\t\rangle &= \t_i (-1)^{\t_1+\cdots+\t_{i-1}}  |\t_1,\ldots,\t_{i-1},\t_i-1,\t_{i+1},\ldots\rangle, \label{Fi-}\\
 F_i^+ |\t\rangle &= (1-\t_i) (-1)^{\t_1+\cdots+\t_{i-1}} |\t_1,\ldots,\t_{i-1},\t_i+1,\t_{i+1},\ldots\rangle. \label{Fi+}
\end{align}
When the system is in the state $|\t\rangle$, the ``number of particles'' is equal to $|\t|=\sum_i \t_i$.

Here we introduce a generalization of the system of fermions, by requiring that the total number of particles has an upper bound~$p>0$, where $p\in\N$. 
When $|I|=\infty$, $p$ can be any positive integer; when $|I|=n$, the generalization makes sense when $p<n$.
Since the total number of particles in the system is relevant, it is logical to introduce also a ``total number operator'' $N$.
So, for the new system, we consider a set of operators $f_i^\pm$ ($i\in I$), together with a number operator $N$. 
We shall sometimes refer to ${\cal F}$ as the algebra generated by the elements $f_i^\pm$ and $N$, subject to the relations that follow hereafter, \eqref{fpfp}-\eqref{fmfp}.
The new operators are determined by their (anti-)commutation relations. 
Explicitly, \eqref{FpFp} is conserved:
\begin{equation}
 \{ f_i^+,f_j^+\} = \{ f_i^-,f_j^-\} = 0. 
\label{fpfp}
\end{equation}
Since $f_i^\pm$ are still interpreted as creation and annihilation operators, the relation with $N$ is determined by
\begin{equation}
[N, f_i^\pm] = \pm f_i^\pm.
\label{Nf}
\end{equation}
The main difference is the generalization of \eqref{FmFp}, which is postulated as follows:
\begin{equation}
(1-\frac{N-1}{p}) f_i^- f_j^+ + (1-\frac{N}{p}) f_j^+ f_i^- = (1-\frac{N}{p})(1-\frac{N-1}{p})\delta_{ij}, \qquad (i,j\in I). 
\label{fmfp}
\end{equation}
In other words, fractional coefficients are introduced in the anti-commutator relation~\eqref{FmFp}. 
It will soon be clear why the anti-commutator between a creation and annihilation operator takes this form.
Certainly, it is already clear that (when $I=\N$) in the limit $p \rightarrow\infty$ this goes to~\eqref{FmFp}.
Formally, we have
\begin{defi}
Let $p$ be a positive integer.
${\cal F}$ is the unital associative algebra generated by the elements $f_i^\pm$ ($i\in I$) and $N$ subject to the relations
\begin{align*}
& \{ f_i^+,f_j^+\} = \{ f_i^-,f_j^-\} = 0, \qquad [N, f_i^\pm] = \pm f_i^\pm,\\
& (1-\frac{N-1}{p}) f_i^- f_j^+ + (1-\frac{N}{p}) f_j^+ f_i^- = (1-\frac{N}{p})(1-\frac{N-1}{p})\delta_{ij}, \qquad (i,j\in I). 
\end{align*}
Moreover, there is a star relation (anti-involution) of ${\cal F}$ determined by
\begin{equation}
(f_i^\pm)^\star = f_i^\mp, \qquad N^\star=N.
\label{star}
\end{equation}
\end{defi}

The important subject of this paper is the Fock space representation $W(p)$ for this algebra. 
It is natural to require that $W(p)$ is generated by a vacuum vector $|0\rangle$ and that the following actions hold:
\begin{equation}
f_i^- |0\rangle =0\qquad (i\in I),  \qquad N|0\rangle = 0.
\end{equation}
Observe that we do not use a different notation for the representatives of $f_i^\pm$ and $N$ in the representation space, compared to the abstract generators, since the meaning is clear in what follows.
The Fock space should also have a bilinear form $\langle\cdot|\cdot\rangle$ which is an inner product on $W(p)$. 
For this form, we require $\langle 0 | 0\rangle =1$ and the hermiticity conditions $(f_i^\pm)^\dagger = f_i^\mp$, $N^\dagger=N$, in agreement with the star relations~\eqref{star}.
These requirements are sufficient to determine the form $\langle\cdot|\cdot\rangle$ on $W(p)$.
Finally, a further natural condition is that $W(p)$ should be an irreducible representation of ${\cal F}$.
\begin{defi}
The Fock space $W(p)$ is an irreducible representation of ${\cal F}$, generated by a vacuum vector $|0\rangle$ and determined by 
\begin{equation}
f_i^- |0\rangle =0\quad (i\in I);  \qquad N|0\rangle = 0;\qquad \langle 0 | 0\rangle =1;\qquad (f_i^\pm)^\dagger = f_i^\mp,\quad N^\dagger=N.
\label{fock}
\end{equation}
\end{defi}
Observe that this is the same definition as the Fock space for ordinary fermions.
This is a major difference with the parastatistics approach, 
where the introduction of a Fock space requires additional relations such as $f_i^- f_j^+ |0\rangle = \delta_{ij} p |0\rangle$~\cite{Palev2003}.

We are now in a position to examine a set of basis vectors of $W(p)$.
Acting on $|0\rangle$ by creation operators, and using~\eqref{fpfp}, yields new vectors of the Fock space $W(p)$ of the following form
\begin{equation}
f^+_{i_1} f^+_{i_2} \ldots f^+_{i_k} |0\rangle,
\label{fff0}
\end{equation}
where $i_1$, $i_2$, $\ldots$, $i_k$ should be mutually distinct. 
In fact, again by~\eqref{fpfp}, it is sufficient to consider vectors of the form~\eqref{fff0} with $i_1 < i_2 < \cdots < i_k$.
Since by~\eqref{Nf}, $[N, f^+_{i_1} f^+_{i_2} \ldots f^+_{i_k}]= k f^+_{i_1} f^+_{i_2} \ldots f^+_{i_k}$, we have
\begin{equation}
N f^+_{i_1} f^+_{i_2} \ldots f^+_{i_k} |0\rangle = k\, f^+_{i_1} f^+_{i_2} \ldots f^+_{i_k} |0\rangle.
\end{equation}
Since the algebra of generalized fermion operators is generated by $f^\pm_i$ and $N$, we should also consider the action of (products of) $f^-_j$ on~\eqref{fff0}.
First of all, consider the action of~\eqref{fmfp} (with $i$ and $j$ interchanged) on a vector~\eqref{fff0} with $k=p$. This yields:
\begin{equation}
f_j^- f_i^+ f^+_{i_1} f^+_{i_2} \ldots f^+_{i_p} |0\rangle =0;
\end{equation}
otherwise said, a state~\eqref{fff0} with $k=p+1$ is annihilated by all annihilation operators $f_j^-$. 
As a consequence, all vectors~\eqref{fff0} with $k>p$ belong to an invariant submodule of the ${\cal F}$-module generated by $|0\rangle$. 
Since the Fock space $W(p)$ is required to be an irreducible representation, all vectors~\eqref{fff0} with $k>p$ are zero in the Fock space (which is strictly speaking a quotient module where vectors of the invariant submodule are factored out).
Hence only vectors~\eqref{fff0} with $k\leq p$ belong to $W(p)$. 
Essentially, this is the argument that motivates the generalization of~\eqref{FmFp} to~\eqref{fmfp}.

Next, let us show that all vectors~\eqref{fff0} with $i_1 < i_2 < \cdots < i_k$ form a basis for $W(p)$.
We shall do this by computing the action of all generators of ${\cal F}$ on these vectors.
As before, the following notation will be useful:
\begin{equation}
|\t\rangle = |\t_1,\t_2,\ldots\rangle= (f_1^+)^{\t_1} (f_2^+)^{\t_2} \ldots |0\rangle,
\qquad \t_i\in\{ 0,1\}, \qquad \sum_i\t_i \leq p .
\label{t}
\end{equation}
The action of $f_i^+$ follows from~\eqref{fpfp} and the previously determined restriction $\sum_i\t_i\leq p$:
\begin{align}
f_i^+ |\t\rangle 
&= 0 \qquad\hbox{ if }\sum_i\t_i = p; \nonumber\\
&= (1-\t_i) (-1)^{\t_1+\cdots+\t_{i-1}} |\t_1,\ldots,\t_{i-1},\t_i+1,\t_{i+1},\ldots\rangle, \qquad\hbox{ if }\sum_i\t_i < p.
\label{fi+}
\end{align}
For a vector~\eqref{t} with $\sum_i\t_i=k$, let us show that
\begin{equation}
f_i^- |\t\rangle = \t_i (-1)^{\t_1+\cdots+\t_{i-1}} (\frac{p-k+1}{p}) |\t_1,\ldots,\t_{i-1},\t_i-1,\t_{i+1},\ldots\rangle. \label{fi-}
\end{equation}
We will prove this by induction on~$k$. 
For $k=0$, all $\t_i$ are zero, and hence~\eqref{fi-} holds.
Acting by~\eqref{fmfp} on the vacuum vector gives
\begin{equation}
(1+\frac{1}{p})f_i^- f_j^+ |0\rangle = (1+\frac{1}{p})\delta_{ij} |0\rangle,
\end{equation}
hence~\eqref{fi-} is also valid for $k=1$.
Let us now assume that it is valid for a value $k=m (<p)$ (induction hypothesis). 
The action of~\eqref{fmfp} with $i=j$ on a vector $|\t\rangle$ with $\sum_i\t_i=m$ gives:
\begin{equation}
(1-\frac{m-1}{p}) f_j^-f_j^+ |\t\rangle +(1-\frac{m}{p}) f_j^+ f_j^- |\t\rangle = (1-\frac{m}{p})(1-\frac{m-1}{p}) |\t\rangle.
\label{tmp}
\end{equation}
The first term herein is zero when $\t_j=1$ in $|\t\rangle$, and then~\eqref{tmp} is valid because of the induction hypothesis, but yields nothing new.
So let us now take $\t_j=0$ in $|\t\rangle$. 
This gives 
\[
(1-\frac{m-1}{p}) (-1)^{\t_1+\cdots+\t_{j-1}}f_j^- |\t_1,\ldots,\t_{j-1},\t_j+1,\t_{j+1},\ldots\rangle + 0
 = (1-\frac{m}{p})(1-\frac{m-1}{p}) |\t\rangle,
\]
or
\begin{equation}
 f_j^- |\t_1,\ldots,\t_{j-1},\t_j+1,\t_{j+1},\ldots\rangle 
 = (-1)^{\t_1+\cdots+\t_{j-1}}(\frac{p-m}{p}) |\t\rangle,
\end{equation}
which implies that~\eqref{fi-} is also valid for $k=m+1$ and $\t_i=1$.
To see that~\eqref{fi-} holds for $k=m+1$ and $\t_i=0$, observe that
\begin{align}
(1-\frac{m-1}{p})f_i^- |\t\rangle &= (1-\frac{N-1}{p})f_i^- f_{i_1}^+ f_{i_2}^+ \ldots f_{i_{m+1}}^+ |0\rangle \nonumber\\
& = -(1-\frac{N}{p}) f_{i_1}^+ f_i^- f_{i_2}^+ \ldots f_{i_{m+1}}^+ |0\rangle = 0.
\end{align}
Herein, we used $i\not\in\{i_1,\ldots,i_{m+1}\}$ (since $\t_i=0$), \eqref{fmfp} and finally the induction hypothesis.

Finally, the requirement $\langle 0|0\rangle$ and the hermiticity condition induce a form on $W(p)$, which is an inner product.
It is easy to compute the inner product for the basis vectors:
\begin{equation}
\langle \t|\t\rangle= \frac{p!}{p^k(p-k)!}, \qquad \sum_i\t_i=k.
\end{equation}
This is again proven by induction on $k$.
For $k=0$ it is clear. 
Consider now a vector $|\t\rangle$ with $\sum_i\t_i=k+1$. 
Then
\begin{align}
\langle \t|\t\rangle &=\langle 0| f^-_{i_{k+1}} \ldots f^-_{i_2} f^-_{i_1} f^+_{i_1} f^+_{i_2} \ldots f^+_{i_{k+1}}|0\rangle \nonumber\\
&= (\frac{p-k}{p}) \langle 0| f^-_{i_{k+1}} \ldots f^-_{i_2} f^+_{i_2} \ldots f^+_{i_{k+1}}|0\rangle = (\frac{p-k}{p}) \frac{p!}{p^k(p-k)!}= \frac{p!}{p^{k+1}(p-k-1)!},
\end{align}
where we used the action of $f^-_{i_1} f^+_{i_1}$ on the vector following on the right, and induction on $k$.

Before summarizing all the above results in a Theorem, let us fix the following notation.
For a vector of the form~\eqref{t}, we shall write $|\t|=\sum_i \t_i$.
Using the above normalization, let us denote
\begin{equation}
|\t\rangle\!\rangle = \sqrt{\frac{p^{|\t|}(p-|\t|)!}{p!}}\, |\t\rangle = \sqrt{\frac{p^{|\t|}(p-|\t|)!}{p!}} (f_1^+)^{\t_1} (f_2^+)^{\t_2} \ldots |0\rangle,
\qquad \t_i\in\{ 0,1\}, \qquad |\t|\leq p .
\end{equation}

\begin{theo}
Let {\cal F} be the unital associative algebra generated by the elements $N$ and $f_i^\pm$ ($i\in I$), subject to the relations~\eqref{fpfp}-\eqref{fmfp}, where $p$ is a positive integer.
Let $W(p)$ be the irreducible Fock space representation of ${\cal F}$, determined by~\eqref{fock}.
Then an orthonormal basis of $W(p)$ is given by the vectors $|\t\rangle\!\rangle$ with $\t_i\in\{0,1\}$ and $|\t|\leq p$. 
The action of the generators on $|\t\rangle\!\rangle$ is given by:
\begin{align}
& N |\t\rangle\!\rangle = |\t|\, |\t\rangle\!\rangle,\nonumber\\
& f_i^- |\t\rangle\!\rangle = \t_i(-1)^{\t_1+\cdots+\t_{i-1}} \sqrt{(p-|\t|+1)/p}\, |\t_1,\ldots,\t_{i-1},\t_i-1,\t_{i+1},\ldots \rangle\!\rangle,\label{actions}\\
& f_i^+ |\t\rangle\!\rangle = (1-\t_i)(-1)^{\t_1+\cdots+\t_{i-1}} \sqrt{(p-|\t|)/p}\, |\t_1,\ldots,\t_{i-1},\t_i+1,\t_{i+1},\ldots \rangle\!\rangle. \nonumber
\end{align}
\label{theo-f}
\end{theo}

Clearly, for the generalized fermions, since $\t_i\in\{0,1\}$ each orbital is occupied by at most one particle (exclusion principle).
Furthermore, the total number of particles in the system is restricted by~$p$ since $|\t|\leq p$.
So this system satisfies the required properties proposed in this paper.

Also from these equations it is clear that in the limit $p\rightarrow\infty$, the action tends to~\eqref{Fi-}-\eqref{Fi+}.
Up to an overall constant, the above actions coincide with the ones from $A$-superstatistics~\cite{Palev2003}, 
which is why we can identify a group theoretical framework for the operators $f_i^\pm$, to be elaborated in the following section.

\setcounter{equation}{0}
\section{Group theoretical interpretation of generalized fermion operators} 
\label{sec:C}%

Although the algebra ${\cal F}$ determined by its generators and relations has no specific further structure, its representation in the Fock space does have an interesting structure.
To see this, let
\begin{equation}
e_{ij}=p\,\{ f_i^+, f_j^-\}, \qquad (i,j\in I).
\label{eij}
\end{equation}
Then the actions~\eqref{actions} yield:
\begin{align}
e_{ij}|\t\rangle\!\rangle &= \t_j(1-\t_i) (-1)^{\t_i+\cdots+\t_{j-1}}\, | \ldots,\t_i+1,\ldots,\t_j-1,\ldots\rangle\!\rangle, \quad (i<j) \nonumber\\
e_{ii}|\t\rangle\!\rangle &= (p-|\t|+\t_i)\, |\t\rangle\!\rangle, \label{eij-action}\\
e_{ij}|\t\rangle\!\rangle &= -\t_j(1-\t_i) (-1)^{\t_j+\cdots+\t_{i-1}}\, | \ldots,\t_j-1,\ldots,\t_i+1,\ldots\rangle\!\rangle. \quad (i>j) \nonumber
\end{align}
In all of these calculations one has to use the fact that $\t_i\in\{0,1\}$ and thus $\t_i^2=\t_i$.
Next, one can compute the action of $[e_{ij}, f_k^+]$, using~\eqref{actions} and~\eqref{eij-action}.
A case-by-case computation ($i=j$ or $i\ne j$; and the location of $k$ relative to $i$ and $j$) leads to
\begin{equation}
[ e_{ij}, f_k^+ ] \,|\t\rangle\!\rangle = (\delta_{jk} f_i^+ - \delta_{ij} f_k^+) |\t\rangle\!\rangle.
\label{fff+}
\end{equation}
Similarly (or using the hermiticity in the Fock space) one finds
\begin{equation}
[ e_{ij}, f_k^- ] = -\delta_{ik} f_j^- + \delta_{ij} f_k^-,
\label{fff-}
\end{equation}
where one should keep in mind that this relation holds when acting in $W(p)$ and not as a relation between the abstract generators in ${\cal F}$. 
Now, one can compute commutators of $e_{ij}$ when acting in $W(p)$:   
\begin{align}
[ e_{ij}, e_{kl} ] &= [e_{ij}, p(f^+_k f^-_l + f^-_l f^+_k) ]\nonumber\\
&= p  ([e_{ij}, f^+_k] f^-_l +  f^+_k[e_{ij},  f^-_l] + [e_{ij},  f^-_l]f^+_k + f^-_l[e_{ij}, f^+_k]) \nonumber\\
&= \delta_{jk}\; p (f_i^+ f_l^- + f_l^- f_i^+) -\delta_{il}\; p (f_k^+ f_j^- + f_j^- f_k^+)\nonumber\\
&= \delta_{jk} e_{il} - \delta_{il} e_{kj}. 
\label{ee}
\end{align}

The relations holding for the operators $e_{ij}$ are the familiar operators of a Lie algebra of type $A$.
More particular, when $I=\{1,\ldots,n\}$ (which will be assumed for the rest of this section), 
these are the standard relations of the Lie algebra $\gl(n)$.
The operators $f_i^\pm$ do not belong to $\gl(n)$.
But it is easy to see that they can be identified with the odd basis elements of the Lie superalgebra $\gl(1|n)$. 
Indeed, let $\gl(1|n)$ be the general linear Lie superalgebra, with standard basis elements $E_{ij}$ ($i,j=0,1,\ldots,n$). 
An element $E_{ij}$ can be identified with an $(n+1)\times(n+1)$-matrix consisting of an entry 1 at position $(i,j)$ and zeros elsewhere.
The odd elements are $E_{0,i}$ and $E_{i,0}$ ($i=1,\ldots,n$) with degree~$\bar 1$, the other elements are even with degree~$\bar 0$.
The bracket in $\gl(1|n)$ is given by
\begin{equation}
\llbracket E_{ij}, E_{kl} \rrbracket = \delta_{jk} E_{il} - (-1)^{\deg(E_{ij})\deg(E_{kl})}\delta_{il} E_{kj}.
\label{gl1n}
\end{equation}
It is now easy to check that with the identification
\begin{align}
&f_i^+ = \frac{1}{\sqrt{p}} E_{i0}, \qquad f_i^-=\frac{1}{\sqrt{p}}E_{0i} \qquad (i=1,\ldots,n), \nonumber\\
& e_{ij}= E_{ij} \quad(i\ne j), \qquad e_{ii}=E_{00}+E_{ii} \qquad (i,j=1,\ldots,n),
\label{fE}
\end{align}
all relations~\eqref{eij}, \eqref{fff+}, \eqref{fff-} and~\eqref{ee} are in agreement with~\eqref{gl1n}, as operators in the Fock space $W(p)$.
In fact, the algebra generated by~\eqref{fE} is the simple Lie superalgebra $\ssl(1|n)$.
The number operator $N$ in this representation belongs to $\gl(1|n)$ and can be written as:
\begin{equation}
N=p - E_{00}=  \sum_{i=1}^n E_{ii},
\label{NE}
\end{equation}
since in this representation the identity operator can be written as $\frac{1}{p} (E_{00}+\sum_{i=1}^n E_{ii})$.

Next, we should identify the Fock space $W(p)$ with an irreducible representation of $\gl(1|n)$.
With the current identification, the creation operators $f_i^+$ are negative root vectors and the annihilation operators $f_i^-$ are positive root vectors of $\gl(1|n)$.
Then it is clear that the vacuum is a highest weight vector of $W(p)$.
The following action is consistent with~\eqref{eij-action} and~\eqref{fE}:
\begin{equation}
E_{00} |0\rangle\!\rangle = p |0\rangle\!\rangle, \qquad E_{ii} |0\rangle\!\rangle = 0 \qquad (i=1,\ldots,n).
\end{equation}
Hence, in the classical $\epsilon$-$\delta$-basis of the $\gl(1|n)$ weight space, 
the highest weight $\Lambda$ of $W(p)$ is given by $\Lambda = p \epsilon+\sum_{i=1}^n 0 \delta_i$, or in coordinates:
\begin{equation}
\Lambda= (p; 0,0,\ldots,0).
\end{equation}
This is the highest weight of a covariant $\gl(1|n)$ representation, so its decomposition with respect to the subalgebra $\gl(1)\oplus\gl(n)$ is well known~\cite{HKTV}.
In terms of highest weights, this is
\begin{equation}
(p; 0,0,\ldots,0) \rightarrow (p)(0,\ldots,0) \oplus (p-1)(1,0,\ldots,0) \oplus (p-2)(1,1,0,\ldots) \oplus \cdots \oplus (0)(1,\ldots,1,0\ldots,0),
\label{branching}
\end{equation}
where $p$ 1's appear in the last contribution.
The characters and dimensions of these representations are well known.
The dimensions are:
\begin{equation}
1 + n + \binom{n}{2} + \binom{n}{3} +\cdots+ \binom{n}{p},
\end{equation}
in agreement with the labeling of the basis vectors $|\t\rangle\!\rangle$ with $|\t|\leq p$.

We can summarize the results of this section as follows.
\begin{theo}
In the Fock space $W(p)$ the representatives of the generators $f_i^\pm$ and $N$ of the algebra ${\cal F}$ (with $I=\{1,\ldots,n\}$) satisfy the bracket relations of the Lie superalgebra $\gl(1|n)$.
$W(p)$ is the irreducible covariant representation of $\gl(1|n)$ with highest weight $(p;0,\ldots,0)$.
\end{theo}

Now that we have identified our generalized fermion operators and their Fock space with the ones arising in 
$A$-superstatistics,
we can refer to~\cite{Palev2003} for an initial study of the corresponding quantum statistics.
For example, it is clear that the so-called grand partition function follows from the character of~\eqref{branching}, 
and all macroscopic properties are determined by this grand partition function. 

\setcounter{equation}{0}
\section{Generalized boson operators} 
\label{sec:D}%

The purpose of this section is to present a generalization of boson operators, in a similar way as we did for fermion operators.
The setup of this section is close to that of section~\ref{sec:B}.
Consider a set of ordinary boson creation and annihilation operators $B_i^\pm$, where $i$ belongs to an index set $I$. 
The defining relations are:
\begin{align}
& [ B_i^+,B_j^+] = [ B_i^-,B_j^-] = 0, \label{BpBp}\\
& [ B_i^-,B_j^+] = \delta_{ij}, \qquad (i,j\in I). \label{BmBp}
\end{align}
The Fock space is generated by a vacuum vector $|0\rangle$ and determined by
\begin{equation}
B_i^- |0\rangle =0\qquad (i\in I),  \qquad \langle 0 | 0\rangle = 1
\end{equation}
and the hermiticity conditions $(B_i^\pm)^\dagger = B_i^\mp$.
In this case, the orthonormal basis vectors of the Fock space can be written as
\begin{equation}
|l\rangle = |l_1,l_2,\ldots\rangle= \frac{1}{\sqrt{l_1!l_2!\cdots}}(B_1^+)^{l_1} (B_2^+)^{l_2} \ldots |0\rangle,
\qquad l_i\in\N,
\end{equation}
and the space is infinite-dimensional.
The action of the creation and annihilation operators on the above vectors is given by
\begin{align}
 B_i^- |l\rangle &= \sqrt{l_i} \, |l_1,\ldots,l_{i-1},l_i-1,l_{i+1},\ldots\rangle, \label{Bi-}\\
 B_i^+ |l\rangle &= \sqrt{l_i+1} \, |l_1,\ldots,l_{i-1},l_i+1,l_{i+1},\ldots\rangle. \label{Bi+}
\end{align}
The ``number of particles'' in the state $|l\rangle$,  is equal to $|l|=\sum_i l_i$.

As before, we introduce a generalization of the system by requiring that the total number of particles has an upper bound~$p>0$, where $p\in\N$. 
The new system is generated by a set of operators $b_i^\pm$ ($i\in I$), together with a number operator $N$. 
Formally, ${\cal B}$ is the algebra generated by the elements $b_i^\pm$ and $N$, subject to similar relations as before.
Explicitly, \eqref{BpBp} is conserved:
\begin{equation}
 [ b_i^+,b_j^+] = [ b_i^-,b_j^-] = 0. 
\label{bpbp}
\end{equation}
As $b_i^\pm$ are still interpreted as creation and annihilation operators, the relation with $N$ reads
\begin{equation}
[N, b_i^\pm] = \pm b_i^\pm.
\label{Nb}
\end{equation}
The main difference is again the generalization of \eqref{BmBp}, which is postulated in a similar form as~\eqref{fmfp}:
\begin{equation}
(1-\frac{N-1}{p}) b_i^- b_j^+ - (1-\frac{N}{p}) b_j^+ b_i^- = (1-\frac{N}{p})(1-\frac{N-1}{p})\delta_{ij}, \qquad (i,j\in I). 
\label{bmbp}
\end{equation}
This commutator between a creation and annihilation operator will lead to the restriction aimed for.
Again, it tends to the usual relation for boson operators when $p$ tends to infinity.
\begin{defi}
Let $p$ be a positive integer.
${\cal B}$ is the unital associative algebra generated by the elements $b_i^\pm$ ($i\in I$) and $N$ subject to the relations
\begin{align*}
& [ b_i^+,b_j^+] = [ b_i^-,b_j^-] = 0, \qquad [N, b_i^\pm] = \pm b_i^\pm,\\
& (1-\frac{N-1}{p}) b_i^- b_j^+ - (1-\frac{N}{p}) b_j^+ b_i^- = (1-\frac{N}{p})(1-\frac{N-1}{p})\delta_{ij}, \qquad (i,j\in I). 
\end{align*}
Moreover, there is a star relation (anti-involution) of ${\cal B}$ determined by
\begin{equation}
(b_i^\pm)^\star = b_i^\mp, \qquad N^\star=N.
\label{star-b}
\end{equation}
\end{defi}

Next, we study the Fock space representation $V(p)$ for this algebra, determined by
\begin{equation}
b_i^- |0\rangle =0\qquad (i\in I),  \qquad N|0\rangle = 0.
\end{equation}
The Fock space should have a bilinear form $\langle\cdot|\cdot\rangle$ which is an inner product on $V(p)$. 
We require $\langle 0 | 0\rangle =1$ and the hermiticity conditions $(b_i^\pm)^\dagger = b_i^\mp$, $N^\dagger=N$, in agreement with the star relations~\eqref{star-b}.
This determines the form $\langle\cdot|\cdot\rangle$ on $V(p)$.
Finally, $V(p)$ should be an irreducible representation of ${\cal B}$.
\begin{defi}
The Fock space $V(p)$ is an irreducible representation of ${\cal B}$, generated by a vacuum vector $|0\rangle$ and determined by 
\begin{equation}
b_i^- |0\rangle =0\quad (i\in I);  \qquad N|0\rangle = 0;\qquad \langle 0 | 0\rangle =1;\qquad (b_i^\pm)^\dagger = b_i^\mp,\quad N^\dagger=N.
\label{fock-b}
\end{equation}
\end{defi}

As for the generalized fermions, we now start the construction of a set of basis vectors of $V(p)$.
Acting on $|0\rangle$ by creation operators, and using~\eqref{bpbp}, yields new vectors of the Fock space $V(p)$ of the following form
\begin{equation}
(b^+_1)^{l_1} (b^+_2)^{l_2} \ldots |0\rangle,
\label{bbb0}
\end{equation}
where $l_i\in\N$ and $\sum_i l_i < \infty$.
Since by~\eqref{Nb}, $[N,(b^+_1)^{l_1} (b^+_2)^{l_2} \ldots ]= (l_1+l_2+\cdots) (b^+_1)^{l_1} (b^+_2)^{l_2} \ldots$, we have
\begin{equation}
N (b^+_1)^{l_1} (b^+_2)^{l_2} \ldots |0\rangle = (l_1+l_2+\cdots)\, (b^+_1)^{l_1} (b^+_2)^{l_2} \ldots |0\rangle.
\end{equation} 
Next, we should consider the action of (products of) $b^-_j$ on~\eqref{bbb0}.
The action of~\eqref{bmbp} (with $i$ and $j$ interchanged) on a vector~\eqref{bbb0} with $l_1+l_2+\cdots=p$ yields:
\begin{equation}
b_j^- b_i^+ (b^+_1)^{l_1} (b^+_2)^{l_2} \ldots |0\rangle =0.
\end{equation}
In other words, a state~\eqref{bbb0} with $l_1+l_2+\cdots =p+1$ is annihilated by all annihilation operators $b_j^-$. 
Hence all vectors~\eqref{bbb0} with $l_1+l_2+\cdots>p$ belong to an invariant submodule of the ${\cal B}$-module generated by $|0\rangle$. 
Since the Fock space $V(p)$ is required to be an irreducible representation, all vectors~\eqref{bbb0} with $l_1+l_2+\cdots>p$ are zero (as before, the space is a quotient module where vectors of the invariant submodule are factored out).
Thus, only vectors~\eqref{bbb0} with $l_1+l_2+\cdots \leq p$ belong to $V(p)$. 

Next, we show that all vectors~\eqref{bbb0} with $l_1+l_2+\cdots \leq p$ form a basis for $V(p)$, by computing the action of all generators of ${\cal B}$ on these vectors.
The following notation will be used:
\begin{equation}
|l\rangle = |l_1,l_2,\ldots\rangle= (b_1^+)^{l_1} (b_2^+)^{l_2} \ldots |0\rangle,
\qquad l_i\in\N, \qquad |l|=\sum_i l_i \leq p .
\label{l}
\end{equation}
The action of $b_i^+$ follows from~\eqref{bpbp} and the restriction $|l|\leq p$:
\begin{align}
b_i^+ |l\rangle 
&= 0 \qquad\hbox{ if }|l| = p; \nonumber\\
&= |l_1,\ldots,l_{i-1},l_i+1,l_{i+1},\ldots\rangle, \qquad\hbox{ if } |l| < p.
\label{bi+}
\end{align}
For a vector~\eqref{l} with $|l|=k$, let us show by induction on $k$ that
\begin{equation}
b_i^- |l\rangle =  l_i (\frac{p-k+1}{p})\, |l_1,\ldots,l_{i-1},l_i-1,l_{i+1},\ldots\rangle. \label{bi-}
\end{equation}
For $k=0$, all $l_i$ are zero, and hence~\eqref{bi-} holds.
Acting by~\eqref{bmbp} on the vacuum vector gives
\begin{equation}
(1+\frac{1}{p})b_i^- b_j^+ |0\rangle = (1+\frac{1}{p})\delta_{ij} |0\rangle,
\end{equation}
so~\eqref{bi-} is also valid for $k=1$.
Let us now assume that it is valid for a value $k=m (<p)$. 
The action of~\eqref{bmbp} with $i=j$ on a vector $|l\rangle$ with $|l|=m$ gives:
\begin{equation}
(1-\frac{m-1}{p}) b_j^-b_j^+ |l\rangle = (1-\frac{m}{p}) b_j^+ b_j^- |l\rangle + (1-\frac{m}{p})(1-\frac{m-1}{p}) |l\rangle.
\label{tmp-b}
\end{equation}
Using induction gives $b_j^+b_j^-|l\rangle=l_j(1-\frac{m-1}{p}) |l\rangle$, and thus we get
$b_j^-b_j^+ |l\rangle =(l_j+1)(1-\frac{m}{p})|l\rangle$, implying that~\eqref{bi-} is also valid for $|l|=m+1$.

The requirement $\langle 0|0\rangle$ and the hermiticity condition induce an inner product on $V(p)$.
It is easy to see that:
\begin{equation}
\langle l|l\rangle= \frac{p! l_1!l_2!\ldots}{p^{|l|}(p-|l|)!}.
\end{equation}
This is again proven by induction on $|l|$.
For $|l|=0$ it is clear. Suppose it holds for vectors with $|l|=m$. Consider a vector $|l\rangle = (b_1^+)^{l_1} (b_2^+)^{l_2} \ldots |0\rangle $ with $|l|=m+1$. 
Let $i$ be the smallest index for which $l_i\ne 0$, and write $|l\rangle = b_i^+ (b_i^+)^{l_i-1} (b_{i+1}^+)^{l_{i+1}} (b_{i+2}^+)^{l_{i+2}}\ldots |0\rangle $.
Then
\begin{align}
\langle l|l\rangle &=\langle 0| \ldots (b_{i+1}^-)^{l_{i+1}} (b_i^-)^{l_i-1} b_i^- b_i^+(b_i^+)^{l_i-1} (b_{i+1}^+)^{l_{i+1}} \ldots |0\rangle \nonumber\\
&= l_i(\frac{p-m}{p}) \langle 0| \ldots (b_{i+1}^-)^{l_{i+1}} (b_i^-)^{l_i-1} (b_i^+)^{l_i-1} (b_{i+1}^+)^{l_{i+1}} \ldots |0\rangle \nonumber\\
&= l_i (\frac{p-m}{p}) \frac{p!(l_i-1)!l_{i+1}!l_{i+2}!\ldots}{p^m(p-m)!}= \frac{p!l_i!l_{i+1}!l_{i+2}!\ldots}{p^{m+1}(p-m-1)!},
\end{align}
where we used the action of $b^-_i b^+_i$ on the vector following on the right, and induction on $m$.

For the normalized vectors, we use the following notation:
\begin{equation}
|l\rangle\!\rangle = \sqrt{\frac{p^{|l|}(p-|l|)!}{p!l_1!l_2!\ldots}}\, |l\rangle = \sqrt{\frac{p^{|l|}(p-|l|)!}{p!l_1!l_2!\ldots}}\, (b_1^+)^{l_1} (b_2^+)^{l_2} \ldots |0\rangle,
\qquad l_i\in\N, \qquad |l|\leq p .
\end{equation}
Then we have the analog of Theorem~\ref{theo-f}:
\begin{theo}
Let {\cal B} be the unital associative algebra generated by the elements $N$ and $b_i^\pm$ ($i\in I$), subject to the relations~\eqref{bpbp}-\eqref{bmbp}, where $p$ is a positive integer.
Let $V(p)$ be the irreducible Fock space representation of ${\cal B}$, determined by~\eqref{fock-b}.
Then an orthonormal basis of $V(p)$ is given by the vectors $|l\rangle\!\rangle$ with $l_i\in\N$ and $|l|\leq p$. 
The action of the generators on $|l\rangle\!\rangle$ is given by:
\begin{align}
& N |l\rangle\!\rangle = |l|\, |l\rangle\!\rangle,\nonumber\\
& b_i^- |l\rangle\!\rangle =  \sqrt{l_i(p-|l|+1)/p}\, |l_1,\ldots,l_{i-1},l_i-1,l_{i+1},\ldots \rangle\!\rangle,\label{actions-b}\\
& b_i^+ |l\rangle\!\rangle =  \sqrt{(l_i+1)(p-|l|)/p}\, |l_1,\ldots,l_{i-1},l_i+1,l_{i+1},\ldots \rangle\!\rangle. \nonumber
\end{align}
\label{theo-b}
\end{theo}

We can make the same remarks as for the generalized fermions.
For the generalized bosons, since $l_i\in\N$ there is no restriction for the number of particles on each orbital.
But since $|l|\leq p$, the total number of particles in the system is restricted by~$p$, thus satisfying the required properties.

From these equations it is clear that in the limit $p\rightarrow\infty$, the action tends to~\eqref{Bi-}-\eqref{Bi+}.
The actions~\eqref{actions-b} coincide (up to a factor) with the ones from $A$-statistics~\cite{Jellal},
so we can again identify a group theoretical framework. 

\setcounter{equation}{0}
\section{Group theoretical interpretation of generalized boson operators} 
\label{sec:E}%

The representatives of the generators of the algebra ${\cal B}$ in the Fock space have a Lie algebraic structure.
Let
\begin{equation}
e_{ij}=p\, [ b_i^+, b_j^-], \qquad (i,j\in I).
\label{eij-b}
\end{equation}
Then the actions~\eqref{actions-b} yield:
\begin{align}
e_{ij}|l\rangle\!\rangle &= \sqrt{(l_i+1)l_j}\, | \ldots,l_i+1,\ldots,l_j-1,\ldots\rangle\!\rangle, \quad (i\ne j) \nonumber\\
e_{ii}|l\rangle\!\rangle &= (l_i+|l|-p)\, |l\rangle\!\rangle. \label{eij-b-action}
\end{align}
Then, one computes the action of $[e_{ij}, b_k^+]$, using~\eqref{actions-b} and~\eqref{eij-b-action}.
As before (case-by-case), this leads to
\begin{equation}
[ e_{ij}, b_k^+ ] \,|l\rangle\!\rangle = (\delta_{jk} b_i^+ + \delta_{ij} b_k^+) |l\rangle\!\rangle.
\label{bbb+}
\end{equation}
Similarly (or using the hermiticity in the Fock space) one finds (as actions in $V(p)$):
\begin{equation}
[ e_{ij}, b_k^- ] = -\delta_{ik} b_j^- - \delta_{ij} b_k^-.
\label{bbb-}
\end{equation}
Thus the commutators of $e_{ij}$, when acting in $V(p)$, satisfy:
\begin{align}
[ e_{ij}, e_{kl} ] &= [e_{ij}, p(b^+_k b^-_l - b^-_l b^+_k) ]\nonumber\\
&= p ([e_{ij}, b^+_k] b^-_l +  b^+_k[e_{ij},  b^-_l] - [e_{ij},  b^-_l]b^+_k - b^-_l[e_{ij}, b^+_k]) \nonumber\\
&= \delta_{jk}\: p(b_i^+ b_l^- - b_l^- b_i^+) -\delta_{il}\: p(b_k^+ b_j^- - b_j^- b_k^+)\nonumber\\
&= \delta_{jk} e_{il} - \delta_{il} e_{kj}. 
\label{ee-b}
\end{align}  
For $I=\{1,\ldots,n\}$, these are again the standard relations of the Lie algebra $\gl(n)$.
The operators $b_i^\pm$ do not belong to $\gl(n)$ but to a larger Lie algebra $\gl(1+n)$.
For $\gl(1+n)$, the standard basis elements are again denoted by $E_{ij}$ ($i,j=0,1,\ldots,n$). 
The bracket in $\gl(1+n)$ is given by
\begin{equation}
\llbracket E_{ij}, E_{kl} \rrbracket = \delta_{jk} E_{il} - \delta_{il} E_{kj}.
\label{gl1+n}
\end{equation}
It is now easy to check that with the identification
\begin{align}
&b_i^+ = \frac{1}{\sqrt{p}} E_{i0}, \qquad b_i^-= \frac{1}{\sqrt{p}}E_{0i} \qquad (i=1,\ldots,n), \nonumber\\
& e_{ij}= E_{ij} \quad(i\ne j), \qquad e_{ii}=E_{ii}-E_{00} \qquad (i,j=1,\ldots,n),
\label{bE}
\end{align}
all relations~\eqref{eij-b}, \eqref{bbb+}, \eqref{bbb-} and~\eqref{ee-b} are in agreement with~\eqref{gl1+n}, as operators in the Fock space $V(p)$.
In this representation, the number operator $N$ can be written as:
\begin{equation}
N=p - E_{00}=  \sum_{i=1}^n E_{ii},
\label{NE-b}
\end{equation}
since in this representation the identity operator is also $\frac{1}{p} (E_{00}+\sum_{i=1}^n E_{ii})$.

It is easy to identify the Fock space $V(p)$ with an irreducible representation of $\gl(1+n)$.
The creation operators $b_i^+$ are negative root vectors and the annihilation operators $b_i^-$ are positive root vectors of $\gl(1+n)$,
so the vacuum is a highest weight vector of $V(p)$.
The following action is consistent with~\eqref{eij-b-action} and~\eqref{bE}:
\begin{equation}
E_{00} |0\rangle\!\rangle = p |0\rangle\!\rangle, \qquad E_{ii} |0\rangle\!\rangle = 0 \qquad (i=1,\ldots,n).
\end{equation}
In the classical $\epsilon_i$-basis of the $\gl(1+n)$ weight space, 
the highest weight $\Lambda$ of $V(p)$ is given by $\Lambda = p \epsilon_0+\sum_{i=1}^n 0 \epsilon_i$:
\begin{equation}
\Lambda= (p; 0,0,\ldots,0).
\end{equation}
This is the highest weight of a covariant $\gl(1+n)$ representation, and its branching with respect to the subalgebra $\gl(1)\oplus\gl(n)$ is well known.
In terms of highest weights, this is
\begin{equation}
(p; 0,0,\ldots,0) \rightarrow (p)(0,\ldots,0) \oplus (p-1)(1,0,\ldots,0) \oplus (p-2)(2,0,\ldots,0) \oplus \cdots \oplus (0)(p,0,\ldots,0).
\label{branching-b}
\end{equation}
Of course, characters and dimensions of such symmetric representations are well known.
The dimensions are:
\begin{equation}
1 + n + \binom{n+1}{2} + \binom{n+2}{3} +\cdots+ \binom{n+p-1}{p}.
\end{equation}

\begin{theo}
In the Fock space $V(p)$, the representatives of the generators $b_i^\pm$ and $N$ of the algebra ${\cal B}$ 
(with $I=\{1,\ldots,n\}$) satisfy the bracket relations of the Lie algebra $\gl(1+n)$.
$V(p)$ is the irreducible covariant representation of $\gl(1+n)$ with highest weight $(p;0,\ldots,0)$.
\end{theo}

The identification of generalized boson operators and their Fock space with the ones arising in $A$-statistics, 
implies that the corresponding quantum statistics coincides with that of~\cite{Jellal}.
The main ingredient is again the character of the Fock space, following from~\eqref{branching-b}, and this yields the 
grand partition function and its inclusive macroscopic properties.

\setcounter{equation}{0}
\section{Conclusions} 
\label{sec:F}%

The symmetrization postulate of quantum physics states that physical systems should be symmetric 
in such a way that the exchange of particles does not lead to distinct observations.
This implies that identical particles exhibit only two types of statistics: fermionic or bosonic.
Theoretically, this leads to the common (anti-)commutation relations imposed on the algebra of creation and annihilation operators.
The corresponding Fock spaces for such systems do not imply any restrictions on the total number of particles.

It is reasonable to assume that there should exist physical systems for which the total number of particles is bounded by some upper limit~$p$, where $p$ is a positive integer.
Such systems require a deformation or generalization of the standard (anti-)commutation relations.
If one wants to preserve the symmetrization property, the creation (resp.\ annihilation) operators of such generalized fermions should still anti-commute, 
and those of generalized bosons should still commute.
So the only relation that could be deformed is the one between a creation and an annihilation operator.
This is precisely the approach followed in this paper.
Fractional coefficients, involving the parameter $p$, are introduced in the relation between a creation and an annihilation operator,
in such a way that the required property (the maximum number of particles is equal to $p$) is satisfied.
These relations are sufficiently simple to allow a complete construction of the Fock space.
This Fock space is determined in the standard way: generated by a vacuum vector that is annihilated by all annihilation operators.

This formalism introduce here is straightforward and physically motivated.
The deviation from standard (anti-)commutation relations is minimal.

Once the Fock spaces for generalized fermions of bosons have been constructed in this paper, 
it becomes clear that such spaces have appeared before as representation spaces of paraparticles of type~$A$.
Parafermions of type~$A$ have been introduced in a Lie superalgebraic setting~\cite{Palev2003}: 
in such an approach, the defining relations are triple relations, 
and the vacuum vector of the representation space is determined by multiple relations (not just the annihilation property).
Parabosons of type~$A$ have been introduced in a similar Lie algebraic setting~\cite{Jellal}.
Thus, for the statistical properties of the generalized fermions and bosons introduced in this paper, 
we can simply refer to~\cite{Palev2003,Jellal}.
The main benefit of the current paper is indeed the alternative approach, starting from simple principles and only a minor deformation of a quadratic relation.

The appliance of our generalized boson and fermion operators in physical models is evident, although we do not consider it as part of this paper.
In existing physical boson models, for example, physical operators expressed in terms of boson creation and annihilation operators can simply be expressed in terms of the generalized operators of this paper.
For $q$-deformations of bosons or oscillators, this procedure has been performed in numerous papers, for example~\cite{Bonatsos2,Bonatsos3,Bonatsos4}.
In those papers and others, a Hamiltonian in terms of boson operators is replaced by the same Hamiltonian in terms of $q$-boson operators, and the effect on the energy spectrum is investigated. 
Such analysis can lead to interesting observations and potential applications.

Let us mention here that $q$-generalizations (of boson models or oscillator algebras) have been framed in a more general context by introducing deformation maps~\cite{Feng,Curtright,Polychronakos}.
The generalization introduced in this paper does not fall into this class of deformation maps.
Indeed, deforming maps (in an algebraic context) in principle preserve the dimension of the representation (of the algebra): only the action of the algebraic operators is deformed.
In our case, the dimension of the representation is actually changed by the parameter~$p$.

To finalize, let us briefly illustrate how our generalized boson operators could be used in a toy model.
Consider a 2-boson system with a given Hamiltonian $H$ of the following form:
\begin{equation}
H=B_1^+B_1^- + B_2^+B_2^-,
\end{equation}
where $B_1^\pm$ and $B_2^\pm$ are ordinary boson creation and annihilation operators satisfying~\eqref{BpBp}-\eqref{BmBp}.
The state space is infinite-dimensional, with basis vectors $|l_1,l_2\rangle$, $l_1,l_2\in\N$.
The energy eigenvalues are determined by
\begin{equation}
H |l_1,l_2\rangle = (l_1+l_2) |l_1,l_2\rangle.
\end{equation}
Hence the energy values (in certain units) and their multiplicities are given by 
\begin{equation}
E_n = n, \qquad \hbox{mult}(E_n)=n+1, \;\; (n=0,1,\ldots,).
\label{E1}
\end{equation}
Let us now consider the same Hamiltonian but for a 2-boson system using our generalization:
\begin{equation}
H=b_1^+b_1^- + b_2^+b_2^-,
\end{equation}
where $b_1^\pm$ and $b_2^\pm$ are generalized boson operators satisfying~\eqref{bpbp}-\eqref{bmbp}.
This time, the state space is finite-dimensional, with basis vectors $|l_1,l_2\rangle$, $l_1+l_2\leq p$.
The energy eigenvalues are computed using~\eqref{actions-b}, and yield
\begin{align*}
H |l_1,l_2\rangle &= \left( l_1(1-\frac{l_1+l_2-1}{p})+l_2(1-\frac{l_1+l_2-1}{p})\right) |l_1,l_2\rangle\\
&= \left( (l_1+l_2)-\frac{1}{p}(l_1+l_2)(l_1+l_2-1) \right) |l_1,l_2\rangle.
\end{align*}
So the energy values are shifted (or deformed by the parameter $p$):
\begin{equation}
E_n = n -\frac{1}{p}n(n-1), \qquad \hbox{mult}(E_n)=n+1, \;\; (n=0,1,\ldots,p).
\label{E2}
\end{equation}
In the non-deformed case~\eqref{E1} the energy levels are equidistant, with distance 1 between consecutive levels.
In the deformed case~\eqref{E2} the energy levels are not equidistant, and the distance between level $n$ and level $n+1$ is given by $1-\frac{2n}{p}$. 
Obviously, when $p$ tends to infinity, the second model reduces to the first model.

Although this toy model is too simple to relate it to realistic physical systems, it does illustrate the potential use of our generalized boson operators in various existing boson models which are popular in the physics literature.


\end{document}